# The Pc1 geomagnetic pulsations in light of the Cepheid theory

A.V. Guglielmi[1,*], A.S. Potapov[2,**], F.Z. Feygin[1,***]

[1]*Schmidt Institute of Physics of the Earth RAS, Moscow, Russia*

[2]*Institute of Solar-Terrestrial Physics SB RAS, Irkutsk, Russia*

[*]guglielmi@mail.ru, [**]potapov@iszf.irk.ru, [***]feygin@ifz.ru

## Abstract

It has been 90 years since the discovery of geomagnetic pulsations in the Pc1 range (0.2–5 Hz), widely known as pearls. In the second half of the last century, the concept of pearls as multiple echoes of a wave packet that propagates along the lines of the geomagnetic field, periodically reflecting off the ionosphere at magnetically conjugate points emerged. This paper proposes an alternative interpretation of the pearls. It is assumed that high above the Earth in the narrow equatorial zone of the outer radiation belt there is a pulsed generator of ion-cyclotron waves. The generator excites a discrete sequence of wave packets, which are recorded in the magnetosphere and on the Earth's surface as a series of pearls. The generator is a Q-modulated ion cyclotron resonator with active filling. The presence of opacity domains adjacent to the resonator's end faces is reminiscent of the opacity layer in the atmosphere of a Cepheid. This association was strengthened by the fact that in both cases the formation of opaque layers is associated with the presence in the medium of ions with different charge-to-mass ratios. Based on this association, the idea of a ponderomotive valve arose, periodically changing the width of the opacity domains, thereby forming a periodic sequence of pearls. The ponderomotive valve in pearl theory is analogous to the Eddington valve in Cepheid theory.

## 1. Introduction

Very low-frequency electromagnetic waves fall from outer space onto the Earth's surface. These are called geomagnetic pulsations [Guglielmi, Troitskaya, 1973; Nishida, 1978], as well as ultra-low-frequency electromagnetic waves [Guglielmi, Potapov, 2021]. Our paper is devoted to the origin of one of the types of pulsations in the Pc1 range (0.2–5 Hz). This species is widely known in literature under the name pearl necklace, or simply pearls. Figure 1 shows the dynamic spectrum of pearls. We see a periodic sequence of quasi-monochromatic wave packets.

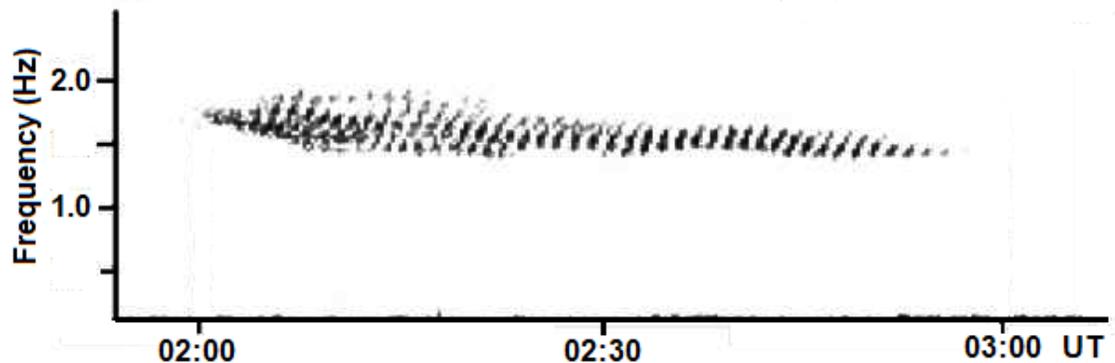

**Fig. 1.** Dynamic spectrum of a series of pearls recorded on March 4, 1986 at the Borok Geophysical Observatory.

The pearls were discovered by Sucksdorff at Sodankylä Observatory and Harang at Tromsø Observatory 90 years ago [Sucksdorff, 1936; Harang, 1936]. In the second half of the last century, the concept of a series of pearls as a multiple echo of a wave packet that propagates in the magnetosphere along the lines of the geomagnetic field, periodically reflecting from the ionosphere at conjugate points developed [Jacobs, Watanabe, 1964; Gendrin, Troitskaya, 1965; Obayashi, 1965; Nishida, 1978; Kangas, Guglielmi, Pokhotelov, 1998].

The pearl echo theory is unsatisfactory in a number of respects. For example, consider the propagation of an Alfvén wave packet in the simplest formulation of the problem. The dispersion equation for Alfvén waves has the form

$$\omega = c_A |k_\parallel|, \qquad (1)$$

where $\omega$ is the wave frequency, $c_A = B/\sqrt{4\pi\rho}$ is the Alfvén velocity, $\rho$ is the plasma density, $k_\parallel$ is the projection of the wave vector **k** onto the external magnetic field **B** [Alfvén, 1950]. The group velocity $\mathbf{v}_g = \partial\omega/\partial\mathbf{k}$ of the wave packet is parallel or antiparallel to the vector **B**. Thus, the trajectory of the packet in the geometric optics approximation coincides with the geomagnetic field line, as assumed in the echo theory.

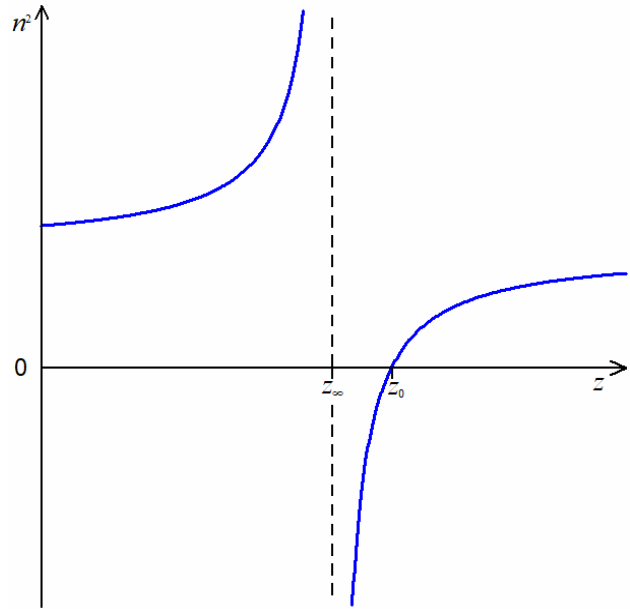

**Fig. 2**. Distribution of the square of the refractive index along the geomagnetic field line in the vicinity of the pole at the gyrofrequency of O⁺ ions. The $z$-axis is directed along the geomagnetic line in the direction of decreasing field.

Let us assume that the wave packet has reached one of the conjugate points in the ionosphere, has been reflected, and is propagating upward in the direction of decreasing geomagnetic field. Now let us take into account that the hydrogen plasma of the magnetosphere contains a small admixture of oxygen ions O⁺. An analysis of the Alfven branch of the dispersion curves shows that the square of the refractive index has a pole at $\omega = \Omega_{O^+}$ and zero at $\omega = \Omega_{O^+}(1+\kappa)$, where $\Omega_{O^+}$ is the gyrofrequency and $\kappa$ is the clarke of oxygen ions. Between zero and the pole there is an opacity band ($n^2 < 0$) [Guglielmi, Troitskaya, 1973]. The width of the opacity band is equal to

$$\Delta z = \left| \frac{1}{B} \frac{\partial B}{\partial z} \right|^{-1} \kappa. \qquad (2)$$

The reflection coefficient from the strip is zero, and the transmission coefficient is

$$D = \exp\left(-\frac{\pi}{2}|k_\parallel|\Delta z\right). \tag{3}$$

At $|k_\parallel|\Delta l > 1$ the wave packet will not reach the conjugate point in the opposite hemisphere.

But even without these and other similar considerations, it is intuitively obvious that the stable discreteness of the pearls clearly indicates that the question of their origin belongs to the class of problems of the nonlinear theory of oscillations, and not at all to the linear theory of wave propagation.

We developed a scenario for exciting pearls in a pulsed ion cyclotron generator located high above the Earth in the narrow equatorial zone of the outer radiation belt [Guglielmi, Potapov, Feygin, 2026a; Potapov, Guglielmi, Feygin, 2026]. The classical theory of Cepheids, known since our student days [Zhevakin, 1953, 1954a,b, 1963], served as an example to follow and a guide for us. However, the search for a mechanism for pearls analogous to the mechanism for Cepheid brightness pulsations proved long and difficult. Using a linear approximation, we constructed a theory of an ion cyclotron resonator (ICR), at the ends of which are opacity stripes analogous to the opacity layer in the atmosphere of a Cepheid [Guglielmi, Potapov, Russell, 2000]. Difficulties arose in searching for a nonlinear mechanism for excitation of a periodic sequence of discrete signals in the ICR. The general idea was that the ponderomotive forces of oscillations excited in the resonator act on the opacity bands, thereby forming a kind of ponderomotive valve, analogous to the Eddington valve in Cepheid theory. When analyzing the action of ponderomotive forces in the magnetosphere, it was necessary to solve a number of methodological problems (see, for example, the review [Lundin, Guglielmi, 2006]). We decomposed the ponderomotive force into the sum of partial ponderomotive forces and analyzed the distribution of ions with different charge-to-mass ratios along the geomagnetic field lines in the static approximation [Guglielmi, Feygin, 2023; Guglielmi, Feygin, Potapov, 2024]. We have considered quantitatively stationary plasma flows [Guglielmi, Potapov, Feygin, 2025; Guglielmi, Potapov, Feygin, 2026a]. To quantitatively describe the operation of a ponderomotive valve, it was necessary to solve the problem of unsteady flow of a complex configuration. We were unable to do this analytically, but knowing the directions and nature of the action of partial ponderomotive forces, we presented a qualitative picture of the pearl mechanism, similar to the kappa mechanism of Cepheids.

The similarity between the mechanisms of such disparate phenomena is interesting from a cognitive standpoint. Furthermore, the emerging analogy strengthens our confidence that we are on the right path to understanding the discrete nature of pearls.

## 2. Pearl Generator

The pearl generator (PG) is designed to consist of three blocks: an ion cyclotron resonator, a pump mechanism, and a ponderomotive Q-modulator [Guglielmi, Potapov, Feygin, 2026b; Potapov, Guglielmi, Feygin, 2026]. The resonator accumulates oscillation energy, continuous pumping of free energy ensures long-term operation of the PG, and Q-modulation of the resonator leads to a pulsed mode of operation of the PG.

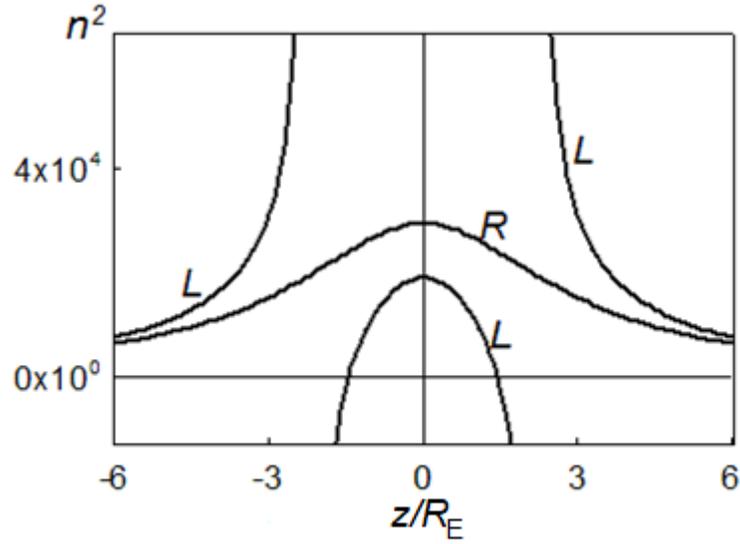

**Fig. 3**. The square of the refractive index of transverse electromagnetic waves of left (*L*) and right (*R*) circular polarization with a frequency equal to 0.1 of the gyrofrequency of H+ ions at $z = 0$. The plasma concentration is 3.12 cm$^{-3}$. The ion composition: H+ (96%), He+ (2%) and O+ (2%). The horizontal axis is the distance along the geomagnetic line intersecting the outer radiation belt. Here $R_E$ is the radius of the Earth. The point $z = 0$ is located at the minimum of the geomagnetic field. The ion cyclotron resonator is located between the zeros of the refractive index for the *L*-wave.

The idea of an ion cyclotron resonator (ICR) was proposed in [Guglielmi, Potapov, Russell, 2000]. The elementary theory of the resonator is developed in the works [Guglielmi, Kangas, Potapov, 2001; Guglielmi, 2007; Klimushkin et al., 2010; Guglielmi, Potapov, 2012, 2021; Mikhailova, 2014; Guglielmi, Potapov, Feygin, 2026b; Potapov, Guglielmi, Feygin, 2026]. Figure 3 illustrates how ICR occurs in magnetospheric plasma containing several types of ions. The resonator is located between the reflection points ($n^2 = 0$). Two opaque bands ($n^2 < 0$) are adjacent to the resonator on the outside. The reflection coefficient from the opaque band is

$$R = 1 - D^2. \tag{4}$$

At $|k_\parallel|\Delta l > 1$, the reflection coefficient is close to unity. The resonator thus has a high quality factor.

Let us idealize reality and assume that the parameters of the environment and the components of the alternating electromagnetic field depend only on $z$. Let us assume for simplicity that the cold magnetospheric plasma contains only two types of ions: light ions (e.g., H$^+$) and a small admixture of heavy ions (e.g., O$^+$ or He$^+$). We will supply indices 1 and 2 to the quantities related to the first and second populations of ions. Let's introduce the following constraints to avoid writing out cumbersome formulas: $m_1 \ll m_2$, $\kappa = \rho_2/\rho \ll 1$, and $\Omega_2 < \omega \ll \Omega_1$ inside the resonator. Here $m_i$ is the mass, $\Omega_i$ is the gyrofrequency. Then, at $R = 1$, the natural frequencies of the ICR are equal to

$$\omega_s = \Omega_2 + \frac{3\sqrt{2\kappa}c_A}{R_E L}\left(s + \frac{1}{2}\right), \quad s = 0, 1, 2, \ldots, \tag{5}$$

where $R_E$ is the radius of the Earth, and $L$ is the McIlwain parameter. The spectrum is equidistant at $s > 0$, but not harmonic. The longitudinal size of the resonator, i.e. the distance between its reflecting ends, increases with the growth of the spectral line number:

$$l_s = 2\left[\frac{\kappa c_A R_E L}{\Omega_2}\left(s + \frac{1}{2}\right)\right]^{1/2}. \tag{6}$$

In reality, the resonator is three-dimensional and open. It is assumed that its lateral "walls" are the caustic surface. The mathematical theory of the three-dimensional open ICR has not been constructed, and in this respect our theory is incomplete.

It is very important that the resonator is a flow-through one. It vaguely resembles a flow cultivator, well-known in biophysics. Energetic protons from the outer radiation belt freely flow into the ICR through the eastern side of the lateral surface.

The protons of the outer radiation belt are in a nonequilibrium state. Their effective transverse temperature $T_\perp$ is greater than the longitudinal temperature $T_\parallel$ [Roederer, 1970]. At $T_\perp > T_\parallel$, ion-cyclotron instability may occur in the plasma [Shafranov, 1963]. In a homogeneous cold hydrogen plasma placed in a uniform external magnetic field and containing a small admixture of energetic protons, the instability increment at low frequencies ($\omega \ll \Omega_1$) is equal to

$$\gamma(k_\parallel) = \frac{\sqrt{\pi}}{2}\left(\frac{T_\perp}{T_\parallel} - 1\right)\frac{\omega_p^2 c_A k_m}{c^2 k_\parallel^2}\exp\left[-\left(\frac{k_m}{k_\parallel}\right)^2\right]. \quad (7)$$

Here $k_\parallel = \omega/c_A$, $\omega_p^2 = 4\pi e^2 N_p / m_p$, $e$ and $m_p$ are the charge and mass of the proton, $N_p$ is the concentration of energetic protons, $k_m = \Omega_p / w_\parallel$, $w_\parallel = \sqrt{2T_\parallel / m_p}$. From (7) it follows that the increment is maximum at frequency

$$\omega = (c_A / w_\parallel)\Omega_p. \quad (8)$$

Cornwall was the first to suggest that the pearls are excited by ion-cyclotron instability of the outer radiation belt [Cornwall, 1965]. Observations indicate that the carrier frequency of pearls corresponds well to formula (8) if the parameters included in it are taken at the point $z = 0$, where the increment (7) is maximum [Nishida, 1978].

Cornwell's theory needs to be supplemented in the following respect. Instability with increment (7) is convective, not absolute [Guglielmi, 1979]. This means that the small initial disturbance does not grow exponentially in time, since it is carried out at the Alfvén velocity from the equatorial zone of the radiation belt. In other words, in a plasma containing one type of ion, the radiation belt would act as an amplifier, but not as a generator of ion-cyclotron waves. But the presence of heavy ions in the magnetosphere turns the amplifier into a generator due to the formation of an ICR, the reflective ends of which provide the necessary feedback. Inside the high-quality ICR, an exponential increase in the energy of ion-cyclotron oscillations with increment (7) occurs due to the energy of protons of the outer radiation belt.

We are, however, not interested in the accumulation of energy of ion-cyclotron oscillations in the closed volume of the ICR. We are interested in the mysterious mechanism of formation of discrete wave packets that leave the ICR and rush towards the Earth's surface in the form of a series of pearls. The following consideration will help us in searching for the discreteness mechanism.

The presence of thin opacity bands adjacent to the ends of the ICR vaguely reminded us of a thin opacity layer in the atmosphere of a Cepheid [Zhevakin, 1953]. This association based on the structural similarity of such disparate objects was reinforced by the fact that in both cases, the formation of opaque layers is associated with the presence of ions with different charge-to-mass ratios in the medium. Thus arose the idea of attempting to supplement the ICR with a constant energy source and an automodulator of the opacity band width to create a pulsed pearl generator.

We will focus on the opacity bands adjacent to the ends of the ICR. To simplify the reasoning, let us assume that both stripes have the same width $\Delta z$. The coefficient of wave transmission through the band $D$, the reflection coefficient $R$, and the quality factor of the resonator depend exponentially on the value of $\Delta z$ (see formulas (3), (4)). In turn, $\Delta z$ linearly depends on the clarke of heavy ions (see (2)).

Our hypothesis is that the ICR experiences periodic Q-modulation under the influence of ponderomotive forces. In the high-Q phase, the oscillation amplitude inside the resonator increases with increment (7). Partial ponderomotive forces acting on heavy ions inside the resonator are directed towards the minimum of the geomagnetic field, i.e. towards the center of the resonator [Guglielmi, Feygin, Potapov, 2024; Guglielmi, Potapov, Feygin, 2025].

As a result, heavy ions are raked from the periphery to the center, and the partial pressure in the resonator drops in the vicinity of its ends. At the boundaries between the resonator and the opacity bands, pressure differences arise, under the influence of which heavy ions flow out of the opacity bands. Clarke $\kappa$ decreases in stripes, and the width of the opacity stripes $\Delta z$ decreases accordingly. The parameter $D$ ($R$) increases (decreases) with time and the resonator exponentially quickly loses quality factor.

In the low-Q phase, oscillation generation is disrupted. The wave energy stored in the high-Q phase flows out. This is how two wave packets are formed, which propagate towards the Earth at the Alfvén velocity.

The high quality factor of the ICR is restored as follows. In the presence of oscillations, the excess heavy ions were retained in the resonator by ponderomotive forces. In the low-Q phase, there are no oscillations and ponderomotive forces are absent. Excess partial pressure inside the resonator leads to a flow of heavy ions outward, into the opacity bands. The width of the bands increases and the high quality factor of the ICR is restored. The generation of ion-cyclotron oscillations begins and the process repeats.

There is one correction that needs to be made to the scenario presented. It is unlikely that the $\Delta z_+ = \Delta z_-$ condition will be met in reality, where the $\pm$ signs refer to the opacity bands in the northern/southern hemisphere. We imposed the $\Delta z_+ = \Delta z_-$ condition in order not to complicate an already complex scenario with reservations and additional explanations. In reality, one of the opacity bands is thinner than the other, for example, $\Delta z_+ < \Delta z_-$. Then, through the thinner northern strip, a flow of all the wave energy accumulated in the resonator will arise. The discrete wave packet will be recorded only in the northern hemisphere.

The resonator has entered a low-Q phase. Here we will make a rather speculative assumption regarding the relaxation of the ICR, leading to the restoration of high Q.

When the bands are filled with heavy ions, the band through which the wave packet left the resonator will ultimately be thicker. In the above case, this will be the northern band. Thus, the subsequent wave packet will reach Earth at the southern conjugate point. In support of our hypothesis, we can say that heavy ions enter the northern, but not the southern, band not only from the resonator but also from the medium adjacent to the opacity band on the outside. The point is as follows.

A wave packet propagating from the ICR to the Earth inevitably crosses the refractive index pole at the point where the carrier frequency of the packet is equal to the gyrofrequency of the heavy ions. At this point, located on the outer boundary of the opacity band, the partial ponderomotive force reaches its maximum and is directed towards the band [Guglielmi, Potapov, Feygin, 2025]. A force impulse is generated, causing heavy ions to flow into the strip.

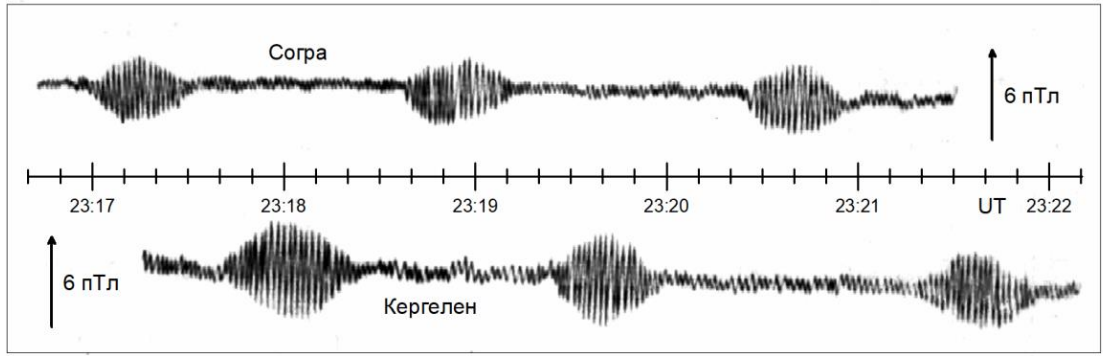

**Fig. 4.** Fragment of the simultaneous registration of pearls at the magnetically coupled stations Sogra and Kerguelen on March 5, 1965.

Thus, pearls appear on the surface of the Earth alternately in the northern and southern hemispheres.

Figure 4 illustrates this effect. It shows pearls registered in the village of Sogra in the Arkhangelsk region and on Kerguelen Island in the Indian Ocean [Gendrin, Troitskaya, 1965; Troitskaya, Guglielmi, 1967].

The presented scenario plausibly describes in simple terms the mechanism of formation of a discrete sequence of pearls, but it is nothing more than "physics in simple terms". A quantitative description could be obtained using a system of electrodynamic and quasi-hydrodynamic equations. However, the non-stationarity of the flow of heavy ions from the opacity bands into the resonator volume and back makes it difficult to find solutions. The nonlinearity of the equations is exponential, which creates additional difficulties.

The scenario we have proposed does not contain an answer to the essential question about the period of ponderomotive modulation of PG, which is intended to coincide with the period of pearl repetition (approximately 2–3 min). Generally speaking, the question of the repetition period should be revisited after our hypothesis about the pearl generator's operating mechanism has been supported by a numerical analysis of the nonlinear system of equations describing the wave field and the nonstationary dynamics of charged particles within the resonator volume and in the opacity bands, as the search for simple solutions encounters analytically insurmountable difficulties. Here, we will limit ourselves to a few general remarks.

Let us recall that within the framework of the echo model, the repetition period is double the travel time of the wave packet along the geomagnetic line from the ionosphere in one hemisphere to the ionosphere in the opposite hemisphere. If we take a naive view and move the conjugate points to the northern and southern ends of the ICR, the repetition period will be equal to

$$\tau = \frac{\sqrt{2}\pi(1+\zeta\kappa)^2}{3(1+\kappa)(1-\zeta)\sqrt{\kappa}} \frac{R_E L}{c_A}, \tag{9}$$

where $\zeta = m_1/m_2$. [Guglielmi et al., 2001]. Calculation using formula (9) gives results that do not contradict observations, but formula (9), obtained within the framework of the linear theory of electromagnetic wave propagation, like the similar formula of the standard echo model, does not solve the problem of the discreteness of pearls. In our opinion, the stable discreteness of a series of pearls is the result of the pulsed generation of wave packets in a substantially nonlinear oscillatory system, and not at all a multiple echo of a single wave packet.

In conclusion of this section of the paper, we will perform a primitive analysis of dimensions in order to roughly estimate $\tau$ by order of magnitude. We have two quantities $\Delta z$ and $l$ with the dimension of length and three quantities $c_A$, $w_2$ and $c_E$ with the dimension of velocity. Here $l$ is the longitudinal size of ICR, $w_2$ is the thermal velocity of heavy ions, $c_E = cE/2B$ is the characteristic velocity of the ponderomotive plasma flow, $E$ is the amplitude of oscillations of the alternating electric field (see, for example, [Guglielmi, Potapov, Feigin, 2026a]). From the given quantities it is possible to form a number of combinations with the dimension of time. It is reasonable to assume that $\tau$ is determined by slow, large-scale processes. This suggests the simplest choice of combinations of $l/w_2$ and $l/c_E$. The estimates based on these formulas are not in sharp contradiction with the measurement data [Guglielmi, Potapov, Feygin, 2026b].

## 3. Discussion

At one time, the pearl echo theory proved useful in that it served as a guideline for planning experimental and theoretical studies of wave phenomena in near-Earth space. It has provided us with images and language for understanding the results of measurements and calculations for decades. We will point out a number of remarkable properties and regularities of Pc1 waves, established through the direct or indirect influence of the fundamental tenets of echo theory.

Apparently, the discovery of an ionospheric MHD waveguide through which pearls propagate along the Earth's surface [Tepley, 1965] suggested the existence of an ionospheric Alfvén resonator (IAR) [Polyakov, 1976; Polyakov, Rapoport, 1981] and stimulated the corresponding observations, which led to the discovery of an amazingly beautiful wave phenomenon in the Pc1 range [Belyaev et al., 1987; Belyaev et al., 1999; Bösinger et al., 2002; Potapov et al., 2014, 2023].

The echo model suggested directions for searching for empirical patterns of Pc1 waves (see review [Kangas et al., 1998]). Thus, in the experiment, a surprisingly stable statistically significant connection $\tau f \approx 10^2$ was established between the repetition period $\tau$ and the carrier frequency $f = \omega/2\pi$ of the pearls. Another significant prediction of the theory, confirmed by experiment, is that the sudden compression of the magnetosphere by the interplanetary shock wave leads to the excitation of a series of pearls in the dayside hemisphere of the magnetosphere. Thus, the interplanetary shock wave can be viewed as a trigger, inducing the transition of the pearl generator from equilibrium to self-excitation. In [Guglielmi et al., 2026], it is hypothesized that the PG belongs to the class of autonomous oscillatory systems with hard self-excitation.

The morphology of Pc1 waves described in the literature based on the results of ground-based and satellite experiments is exceptionally rich. It has been established that pearls are most often excited during the decline of the 11-year cycle of solar activity [Matveeva, Troitskaya, 1965]. Seasonal and daily variations in the frequency of pearl appearance have been carefully studied. It has been found that during a geomagnetic storm, pearls are usually observed in the recovery phase. It has been suggested that during the main phase of a storm, when the geoelectromagnetic field is globally disturbed over a wide range of frequencies, the pearls are simply not visible on magnetograms. A careful spectral-temporal analysis of ULF oscillations in the main phase led to a remarkable discovery [Troitskaya, Melnikova, 1959; Troitskaya, 1961]. It turned out that at the peak of the main phase in the evening sector of the magnetosphere, powerful oscillations in the Pc1 range are excited, radically different from the pearls. The average frequency of the detected oscillations increases monotonically by several octaves over the course of approximately half an hour. Thus, we have a wealth of material at our disposal for analysis within the framework of our proposed interpretation of the pearls.

The variety of alternative hypotheses that a researcher puts forward in the course of calculations or experiments has great heuristic value, regardless of the fact that many preliminary considerations will later be changed or completely discarded. This meaning is that theoretical constructs allow for thought experiments to be carried out, on the basis of which a strategy for searching for new properties and patterns of dynamics of the object under study is selected. We hope that the hypothesis of ponderomotive self-action of ICR oscillations will prove useful in this sense.

On the way to completing the theory of the pearl generator, we encountered an analytically insurmountable obstacle. In the linear approximation, the difficulties arise from the fact that the ICR is a three-dimensional open resonator. In nonlinear theory, an obstacle to finding solutions to dynamic equations is the non-stationarity of the processes of formation of discrete wave packets, as well as the extremely interesting fact that the nonlinearity is exponential.

We presented a qualitative description of pearl excitation and propagation. Based on our description, we can make a number of predictions that are amenable to experimental verification. Let us point out here one trivial prediction of this kind. From the elementary theory of ICR [Guglielmi, Potapov, Russell, 2000] it follows that in a closed volume of the resonator the time-averaged Poynting energy flux is zero. If the oscillations leave the resonator and propagate in the form of a wave packet, then it is quite obvious that the Poynting vector is directed towards the Earth. Apparently, our prediction was confirmed in in situ satellite experiments [Mursula et al., 2001; Loto'aniu et al., 2005].

## 4. Conclusion

The excitation of pearls in the Earth's magnetosphere occurs as a result of a confluence of circumstances. The structure of the pearl generator contains an ion-cyclotron resonator, a mechanism for converting the energy of protons of the outer radiation belt into oscillation energy, and two ponderomotive valves. The generator operates as a result of the self-consistent action of three electrodynamic processes: ion-cyclotron instability of plasma, excitation of standing and traveling waves, and ponderomotive redistribution of heavy ions. A similar confluence of circumstances occurs in the vicinity of other planets in the solar system, as well as in more distant places in outer space. The search for extraterrestrial "pearl necklaces" would be interesting and have a chance of success.

In conclusion, it should be emphasized that the idea of ponderomotive valves was suggested to us by a distant analogy between the periodic sequence of pearls and the periodicity of the luminosity of Cepheids.

*Acknowledgments*. We express our sincere gratitude to B.V. Dovbnya and O.D. Zotov for their long-term collaboration in studying Pc1 geomagnetic pulsations and for their invaluable assistance in preparing this article. This work was supported by the Ministry of Science and Higher Education of the Russian Federation within the framework of state assignments of the Schmidt Institute of Physics of the Earth of the Russian Academy of Sciences and the Institute of Solar-Terrestrial Physics of the Siberian Branch of the Russian Academy of Sciences.